\documentclass[sigconf,screen]{acmart}
\input{utils}

\AtBeginDocument{%
  }

\setcopyright{acmlicensed}
\acmDOI{10.1145/3664646.3664770}
\acmYear{2024}
\copyrightyear{2024}
\acmSubmissionID{fsews24aiwaremain-p64-p}
\acmISBN{979-8-4007-0685-1/24/07}
\acmConference[AIware '24]{Proceedings of the 1st ACM International Conference on AI-Powered Software}{July 15--16, 2024}{Porto de Galinhas, Brazil}
\acmBooktitle{Proceedings of the 1st ACM International Conference on AI-Powered Software (AIware '24), July 15--16, 2024, Porto de Galinhas, Brazil}
\received{2024-04-05}
\received[accepted]{2024-05-04}

\begin{document}

\title[A Case Study of LLM for Automated Vulnerability Repair...]{A Case Study of LLM for Automated Vulnerability Repair: Assessing Impact of Reasoning and Patch Validation Feedback}

\author{Ummay Kulsum}
\orcid{0009-0005-6466-6076}
\affiliation{%
  \institution{North Carolina State University}
  \city{Raleigh}
  \country{USA}
}
\email{ukulsum@ncsu.edu}

\author{Haotian Zhu}
\orcid{0009-0007-2520-8309}
\affiliation{%
  \institution{Singapore Management University}
  \city{Singapore}
  \country{Singapore}
}
\email{htzhu@smu.edu.sg}

\author{Bowen Xu}
\orcid{0000-0002-1006-8493}
\affiliation{%
  \institution{North Carolina State University}
  \city{Raleigh}
  \country{USA}
}
\email{bxu22@ncsu.edu}

\author{Marcelo d'Amorim}
\orcid{0000-0002-1323-8769}
\affiliation{%
  \institution{North Carolina State University}
  \city{Raleigh}
  \country{USA}
}
\email{mdamori@ncsu.edu}








\begin{abstract}
Recent work in automated program repair (APR) proposes the use of \emph{reasoning and patch validation feedback} to reduce the semantic gap between the LLMs and the code under analysis.
The idea has been shown to perform well for general APR, but its effectiveness in other particular contexts remains underexplored.


In this work, we assess the impact of reasoning and patch validation feedback to LLMs in the context of vulnerability repair, an important and challenging task in security. To support the evaluation, we present \tname{}, an LLM-based vulnerability repair technique based on reasoning and patch validation feedback. \tname~(1) uses a chain-of-thought prompt to reason about a vulnerability prior to generating patch candidates and (2)~iteratively refines prompts according to the output of external tools (e.g., compiler, code sanitizers, test suite, etc.) on previously-generated patches.


To evaluate performance, we compare \tname{} against the state-of-the-art vulnerability repair techniques for C and Java using public datasets from the literature. Our results show that \tname{} generates, on average, 14\% and 7.6\% more correct patches than the baseline techniques on C and Java, respectively. We show, through an ablation study, that reasoning and patch validation feedback are critical. We report several lessons from this study and potential directions for advancing LLM-empowered vulnerability repair.
\end{abstract}




\maketitle


\section{Introduction}


\sloppy

\begin{figure*}[ht]
\centering
\includegraphics[width=0.95\textwidth]{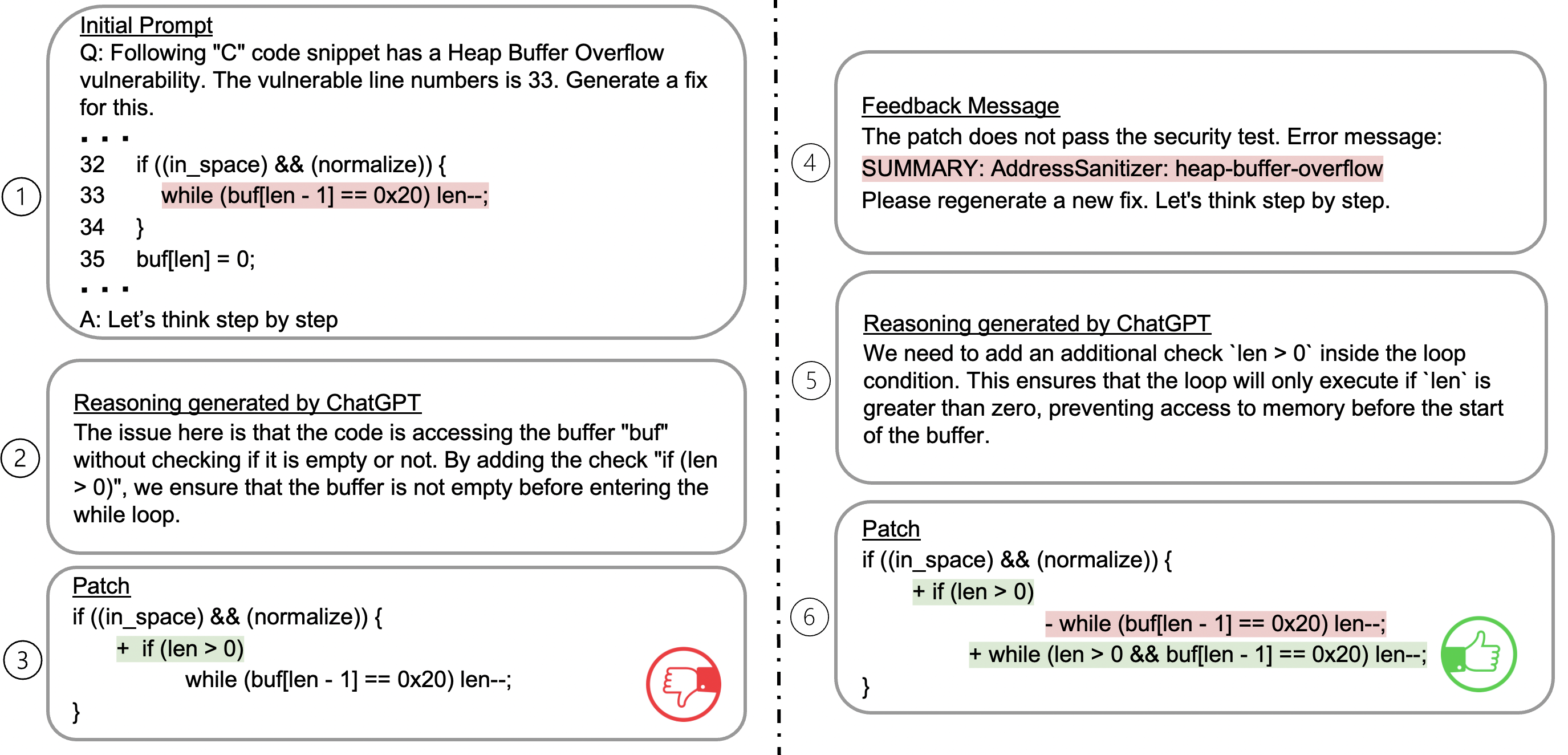}
\vspace{-2mm}
\caption{An example of correct patches generated by \tname{}. 
\textcircled{1} \tname{} queries \chatgpt\ using a prompt with a high-level description  of the vulnerability (``Heap Buffer Overflow'') and a fragment of the vulnerable code. The prompt also includes the trigger ``Let's think step by step'' to access \chatgpt{}'s reasoning feature~\cite{kojima2022large}.
\textcircled{2} In response, \chatgpt\ explains the problem and suggests a solution. \textcircled{3} \tname\ queries the model again combining the initial prompt and reasoning information, and a patch is produced. \textcircled{4} However, this patch does not pass the security tests (\faThumbsODown{}). \tname{} leverages the output of external tools (e.g., compiler, functional and security tests) to circumvent the problem. In this case, \tname{} repeats the previous process after incorporating the error message generated by the address sanitizer (introduced by the compiler). \textcircled{5} \chatgpt\ updates its reasoning and generates another patch. \textcircled{6} This patch passes all tests (\faThumbsOUp{}).} 
\label{fig:intro_example}
\vspace{-2mm}
\end{figure*}

Automated vulnerability repair is an active field of research~\cite{Harer2018LearningTR,gao2021beyond,chen2022neural,zhang2022program,bui2022vul4j,fu2022vulrepair,wu2023effective}. A variety of repair strategies have been proposed in the literature. Recent prior work~\cite{pearce2022examining,wu2023effective} applied large language models (LLMs) for vulnerability repair to mitigate the limitations of traditional non-LLM approaches, such as the inability to adapt to unseen circumstances or the difficulty in obtaining labeled data~\cite{vulnerability-disclosure}.  However, the effectiveness of these approaches is still questionable. We observe that there is an important \emph{semantic gap} between what LLMs know and what they need to know to solve a challenging code-related task, such as vulnerability repair. Intuitively, the LLM knows little about the semantics of the code under analysis.

Recently, Xia et al.~\cite{xia2023keep} proposed~\textsc{ChatRepair}, an LLM-based technique for automated program repair. They demonstrate that \textsc{ChatRepair} can \emph{bridge the semantic gap between the program under analysis and the LLM}. 
\textsc{ChatRepair} automatically repairs bugs by using (1)~reasoning and (2)~patch validation feedback. \emph{Reasoning} enables the LLM to better understand the connection between the task and the code before producing answers whereas \emph{Patch Validation Feedback} enables the LLM to refine its answer based on sensible static and dynamic information (e.g., compilation errors and tests failures due to violation of test case). Intuitively, \textsc{ChatRepair} tries to mimic how developers repair a bug in the real world by first attempting to understand the reasons for the problem (reasoning) and then iteratively refining the answers with the assistance of external tools (feedback).


This paper revisits the \textit{reasoning-feedback} idea that Xia et al. introduced~\cite{xia2023keep} in the context of vulnerability repair, which is a challenging class of problem in code repair. To enable that analysis, we present \tname{}, a LLM-based technique for vulnerability repair based on reasoning and feedback.
In contrast to general code repair, \tname{} brings the kind of vulnerability (e.g., fix a ``Heap Buffer Overflow'' in a fragment of C code)  into the LLM context for better reasoning. Recent prior work~\cite{wei2022chain,kojima2022large,zhang2023getting} have shown that LLMs are capable of performing complex tasks that require intermediate reasoning steps. \tname\ follows a design similar to those approaches to access such ``reasoning'' capability of the LLM.  Complementary to reasoning, patch validation feedback helps the LLM to fix code that does not compile or code that breaks the program functionality~\cite{pinconschi2021comparative}. Intuitively, the output of external tools provides guidance to the LLM remediate those problems~\cite{chatgpt-plugins}. \textbf{\tname\ iteratively incorporates the output of the compiler, test case, and also security sanitizer in prompts} under the assumption that the LLM learns how to avoid them. It is worth noting that Xia et al. focuses on fixing bugs in general and it does not consider security tests.


%


Figure \ref{fig:intro_example} illustrates \tname{} on a running example, consisting of a ``Heap Buffer Overflow'' vulnerability in file \CodeIn{parser.c} from Libxml2, a well-known software library for parsing XML documents.\footnote{\url{https://gitlab.gnome.org/GNOME/libxml2/-/commit/6a36fbe}}
\tname\ starts by requesting an explanation for the problem to the LLM. The answer from the LLM appears in the middle box on the left-hand side of the figure. Then, \tname{} uses that explanation to request a patch to the LLM. The answer from the LLM appears in the box with a ``thumbs down'' icon (\faThumbsODown{}). Unfortunately, although the explanation is coherent with the actual problem, the patch still fails the security test. As such, \tname{} extracts the error message from test execution logs and provides it to the LLM. \tname\ requests the LLM to reason about the problem again. Note from the explanation that, this time, the LLM spots the need to introduce an extra check in the loop conditional. \tname\ uses the revised explanation to request another patch and the LLM produces a plausible patch this time. The answer from the LLM appears in the box with a ``thumbs up'' icon (\faThumbsOUp{}). Indeed, we find that this patch is equivalent to the ground truth, which replaces the original condition in the while statement with the condition \CodeIn{len > 0 \&\& buf[len -1] == 0x20}. 

We conduct three experiments to assess the impact of reasoning and feedback in LLM-based vulnerability repair.

First, we evaluate the effectiveness of \codexvr~\cite{pearce2022examining},\footnote{Pearce and others~\cite{pearce2022examining} did not name their technique. We use the name \codexvr\ for their technique, reflecting the use of the Codex model applied to vulnerability repair.} a recently-proposed technique for LLM-based vulnerability repair. \codexvr\ uses \codex~\cite{openai-codex}, a coding-specific LLM developed by OpenAI, but it is no longer available.
For that reason, we evaluate \codexvr\ with \chatgpt{} and various prompts. This experiment evaluates both robustness and effectiveness of \codexvr.
Based on our results, we find that \codexvr\ with \chatgpt{} is capable of generating more plausible patches than with \codex{} by a large margin, i.e., 20.2\%, showing that the performance of \codexvr\ does not degrade. 
However, we find that, despite the improvement, \codexvr{} only generates a relatively small amount of plausible patches. On average, only \textbf{29.6\%} of the patches it generates are plausible.

Second, we compare \tname{} against an optimized version of \codexvr\ finding that our proposed technique outperforms the baseline. In particular, we find that \tname\ generates, on average, 
\textbf{14\%} and \textbf{7.6\%} more correct patches than the baseline technique on C and Java datasets, respectively. 

Third, we conduct an ablation study to measure the contribution of each component of \tname{}: reasoning and feedback. We find that both components contribute to \tname{}'s performance. For example, considering the percentage of plausible patches produced, we find that the performance of \tname{} falls from \textbf{64\%} to \textbf{34\%} and to \textbf{35\%} when feedback is disabled and when reasoning is disabled, respectively. 
These results show that both components contribute to the performance of \tname\ and the
performance of the full-fledged combination achieves the
best or equally the best performance in 90\% cases.

We learned several lessons from this work. For example, our results indicate that more work is needed to devise automated techniques to mine code contexts to be incorporated in LLM prompts. It is worth noting that some vulnerability repair tasks require domain knowledge (e.g., design choices, environment factors, etc.), posing a challenge in automating this task. It is important to differentiate those cases in benchmarks. As seed for future work, our experience with patch validation suggests that integrating large language models to improve productivity in manual inspection --as opposed to replacing humans altogether-- may offer a good balance between accuracy and scalability.


Our work makes the following contributions:
(1)~We examine the idea of \textit{reasoning-and-feedback} in LLMs for vulnerability repair by presenting \tname{}, a tool follows the spirit of the idea;
(2)~We evaluate \tname{} on datasets of C and Java programs showing that the approach outperforms the state-of-the-art baselines;
(3)~We discuss lessons learned and the implications of our study; 
(4)~We provide the replication package to facilitate future research: \url{http://tinyurl.com/vrpilot-artifacts}.





\section{Methodology}
\label{technique}
\label{sec:overview}

In this section, we first introduce the problem formulation and then detail the design of the method.

\subsection{Problem Formulation}


The primary objective of automated vulnerability repair is to fix software vulnerabilities without human intervention. A repair may be unacceptable because the code fails the \emph{security tests} introduced by the vulnerability detection tool (e.g., ASAN~\cite{asan} for memory safety issues and UBSAN for integer overflows and division by zero~\cite{ubsan}). Likewise, a patch may be unacceptable if the patched code does not pass the \emph{functional tests}.
We define the task of vulnerability repair as follows.



\textbf{Vulnerability Repair (VR)}: Given a vulnerable program $P$, the corresponding security specification \matit{SS} that makes the vulnerable program $P$ fail, the functional specification \matit{FS}, we define VR as a function taking the triple \matit{(P, SS, FS)} as input and producing a patch for $P$ as output. More precisely, the problem of VR is to find a patch $P'$ (i.e., a safe variant of $P$) that passes both \matit{SS} and \matit{FS}. We refer to the patch $P'$ as \emph{plausible patch}.
In the literature, some of the prior works (such as~\cite{Ma2017VuRLEAV}) on VR assume example patches are available. Therefore, they define the example patches as part of the input of the task. In this work, considering VR is often time-critical, we follow a more realistic task definition in this work, \emph{zero-shot setting}, i.e., none of the patch examples is given.

\subsection{Overview} 

\begin{figure}[]
    \centering
    \includegraphics[width=\columnwidth]{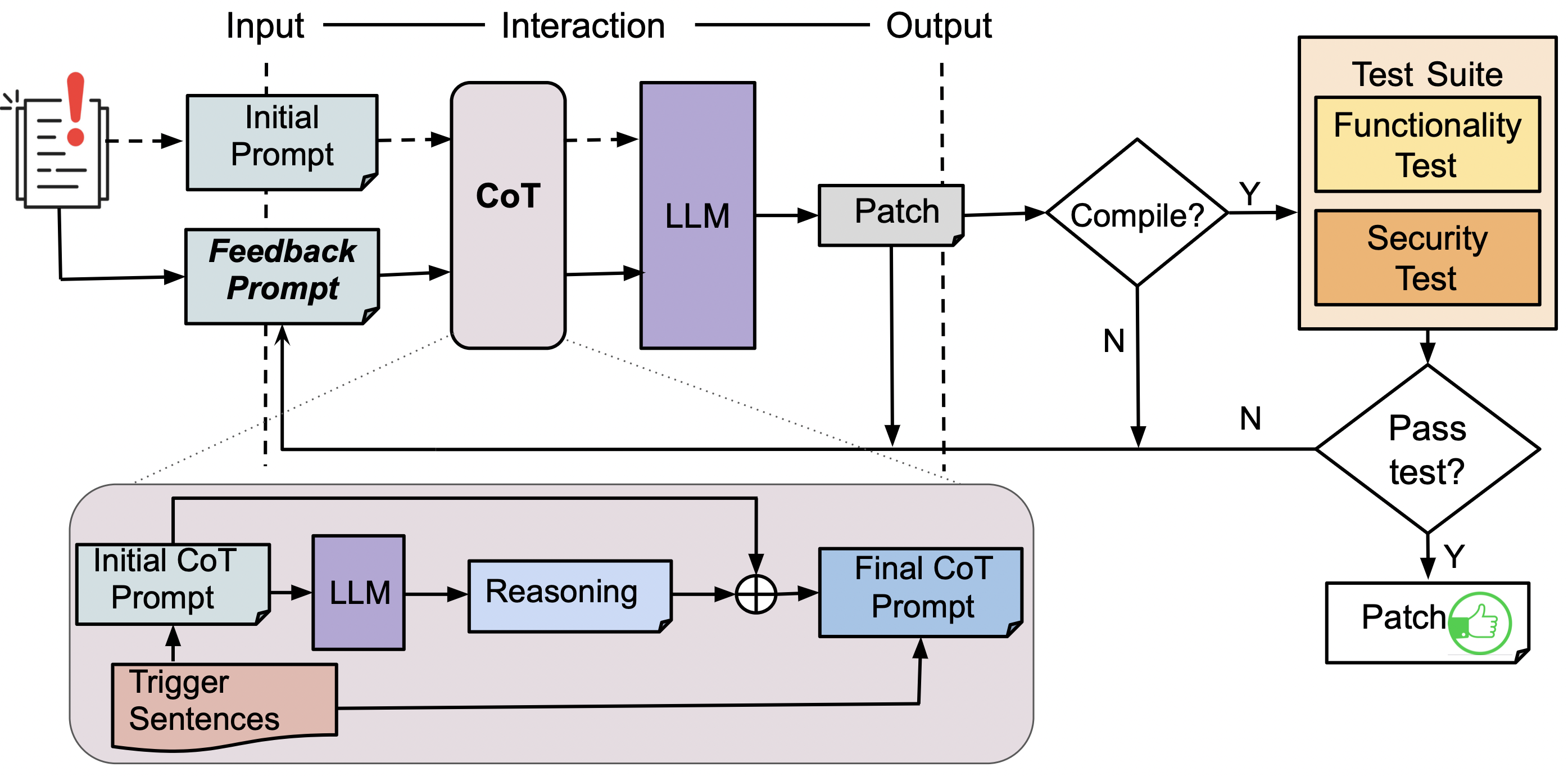}
    \vspace{-5mm}
    \caption{\label{fig:mainDiagram} Overview of \tname{}. The repair process starts by constructing an initial prompt based on vulnerability information and passing it to the Chain-of-Thought block (CoT). The CoT block adds a trigger sentence and queries the LLM to generate reasoning for the task. The final CoT prompt combines the initial CoT prompt, the reasoning, and another trigger sentence. The LLM generates the repair patch using this final CoT prompt, and then the patch is compiled and tested. If the patch passes, it is considered a plausible patch. If not, \tname{} refines the incorrect patch by creating a feedback prompt with the error messages and vulnerability information and repeats the same steps.}
    \vspace{-5mm}
\end{figure}

An LLM (e.g., InCoder~\cite{Fried2022}, Polycoder~\cite{Garousi2013}, and \chatgpt{}~\cite{chatgpt}) is trained on massive amounts of data using general-purpose tasks\footnote{For example, predicting a masked token and predicting the following sentence, which is the approach used in \chatgpt~\cite{chatgpt}.}. Users interact with an LLM through a \emph{prompt}, which describes a task. For Software Engineering tasks, a prompt typically includes code and text~\cite{codexStudy2022}. Users of LLMs often can control the \emph{temperature}, a variable that sets the level of (un)predictability of answers.


\tname{} is an LLM-based vulnerabilty repair technique that employs \emph{reasoning} and (patch validation) \emph{feedback} in producing plausible patches. Conceptually, reasoning enables the LLM to better understand the task that needs to be solved prior to solving the task; it helps an LLM think logically and draw a conclusion based on premises~\cite{huang2022towards} (Section~\ref{subsec:prompt}). Feedback helps an LLM to obtain information from external sources~\cite{chatgpt-plugins}; information that is not present in the models' training data. These two mechanisms contribute to bridging the semantic gap between the LLM and the programs under analysis (Section~\ref{subsec:feedback}).




Figure~\ref{fig:mainDiagram} details the workflow of \tname. The input, highlighted with an exclamation mark, consists of information about the vulnerability, such as the statements in the vulnerable function and description of vulnerability (e.g., ``heap buffer overflow''). The initial prompt includes the description of the vulnerability and the vulnerable function. \tname\ then obtains a patch through a zero-shot chain-of-thought prompting~\cite{kojima2022large}, highlighted in the box at the bottom-left corner of the figure (Section~\ref{subsec:prompt}). If the patch is not compilable, \tname\ updates the prompt with an indication of the compilation error and requests \chatgpt{} for a better solution. Otherwise, \tname\ runs the test suite, including both functional and security tests. 
If a patch passes both tests,
\tname{} reports this plausible patch and the process terminates. Otherwise, \tname{} extracts the useful part of the error message, and updates the prompt accordingly.
\tname{} repeats these steps until reaching the maximum number
of feedback iterations. 

\vspace{-3mm}
\subsection{Chain of Thought} \label{subsec:prompt}
The optimal performance of LLMs relies on the proper design of prompts. Prior studies have shown that various LLMs possess the capacity to generate reasoning and effectively address complex multi-step reasoning tasks with the help of advanced prompt techniques~\cite{wei2022chain,kojima2022large,yao2022react}. For example, Wei et al.~\cite{wei2022chain} demonstrated that providing a series of intermediate reasoning steps explained through a few examples in the prompt increases the performance of LLMs in solving complex mathematical and commonsense tasks, naming that approach as \emph{chain-of-thought prompting}. On followup work, Kojima et al.~\cite{kojima2022large} showed that the success of the chain of thought prompting depends on 
the explanation in a few-shot manner and proposed a new zero-shot approach where the explanations are generated by the LLM. 

Vulnerability repair is undeniably a complex problem that may require intermediate reasoning steps, like comprehending the code, comprehending the vulnerability present in the code, and devising a suitable solution. Consequently, we hypothesize that utilizing LLMs in expressing the rationale for these intermediate steps can boost the performance of LLMs in vulnerability repair.




Kojima et al.~\cite{kojima2022large} proposed a prompt template to be used in solving a given task through zero-shot chain-of-thought (CoT) prompting. It involves two stages of prompting: (1) reasoning extraction and (2) answer extraction. The initial CoT prompt takes the form $X^\prime = Q: [X]$ $ A: [T]$, where $[X]$ is a slot for a problem $X$ and $[T]$ is a sentence that triggers reasoning generation, such as ``Let's think step by step''. The symbols $Q:$ and $A:$ initiating a sentence is a short form of ``Question'' and ``Answer'', respectively. It is assumed that the LLM is trained on completion tasks; as such, it will generate the answer that follows the triggering signal for reasoning T. The final CoT prompt takes the form $[X^\prime] [Z] [A]$, where $[X^\prime]$ is a slot for the  initial CoT prompt, $[Z]$ is for the response generated from the initial CoT prompt, and $[A]$ is for the trigger sentence to extract the solution of the problem.

In our case, we define $X$ as a vulnerability repair task,
the initial prompt contains the vulnerability information, vulnerable code block, and an instruction to repair the vulnerability.
For the trigger sentence $[T]$, we choose to use \emph{``Let's think step by step''} as it achieves the best performance in multiple reasoning task datasets~\cite{kojima2022large}. Then we construct the final prompt by combining the initial prompt, the reasoning and appending a trigger sentence ``Therefore the fixed code is'' to extract the generated repair patch.

\vspace{-3mm}
\subsection{Patch Validation Feedback} \label{subsec:feedback}


\chatgpt{} is trained with reinforcement learning and human feedback to follow conversations. We leverage this observation by iteratively providing previously unused compiler and test suite error messages as conversation feedback. These feedback messages assist \chatgpt{} in refining incorrect patches. Intuitively, error messages are designed to guide developers to fix vulnerabilities. 
\chatgpt{} generates code using the code context provided in the prompt and its internal knowledge. However, it lacks the external knowledge, i.e., security and functionality requirements of the project. \tname\ provides this external information through feedback and guides \chatgpt{} to iteratively improve generated patches. 
Even though incorrect patches might seem like failures, error logs contain valuable information that can help understand the cause of the problem and subsequently modify the patch to fix the vulnerability. To construct comprehensive feedback, we sequentially leverage 3 types of error messages to instruct \chatgpt{} effectively, namely: 


\noindent\textbf{(1) Feedback from compilation test.} When a generated patch does not compile, \tname{} parses the compilation error message and extract the relevant error messages. The extracted relevant error messages are then selected based on their proximity to the vulnerable lines of code within a range of 100 lines. Hence, \tname{} focuses on the specific errors directly related to the problematic code, allowing for more accurate feedback. 

\noindent\textbf{(2) Feedback from functionality test.} \tname{} runs functional tests on the generated patch to ensure it implements the desired functionality. For that reason, \tname{} automatically executes the test suite included in the projects. If the patch fails to pass these functionality tests, \tname{} parses the corresponding log to extract the names of the failed test cases that carry the high-level semantic of the testing purpose. These failed test cases are then used as additional feedback for \chatgpt{}, providing further information about the specific functional deficiencies of the patch.

\noindent\textbf{(3) Feedback from security test.} Similarly, we run security tests by enabling different sanitizer flags (ASAN/UBSAN) according to the original setting of the projects.
The security error message is parsed from the log generated by the test and used as feedback to guide \chatgpt{}.

Based on the above feedback messages, \tname{} constructs the initial CoT prompt for feedback iteration (feedback prompt) with chain-of-thought. We design the feedback prompt as $F = Q:[X][C][E] A:[T]$ where $[C]$ is the slot for code changes suggested by LLM in previous iteration; $[E]$ is the slot for the feedback error message described above; $[X]$ and $[T]$ are slots for the vulnerability repair task and trigger sentence respectively as described in section \ref{subsec:prompt}.
The final CoT prompt for feedback iteration is constructed as $[X^\prime][Z][A]$ where $[X^\prime]$ is a slot for the feedback prompt $F$ and $[Z]$ and $[A]$ are slots for the reasoning and trigger sentence for answer extraction respectively as describe in section \ref{subsec:prompt}. The above iterative process aims to enhance the quality and accuracy of the generated patches by incorporating information from compilation errors, failed functionality tests, and security tests.


\subsection{Implementation} \label{subsec:implementation}
\tname{} is implemented in Python and it accesses \chatgpt{} via API provided by OpenAI~\cite{chatgptapi}. We use \emph{gpt-3.5-turbo} as the underlying model for \chatgpt{}. \tname{} initializes \chatgpt{} with a system message; ``You are a chatbot for vulnerability repair'' for the vulnerability repair task. We iterate through 5 different temperature levels (0.0, 0.25, 0.5, 0.75, 1.0). It is worth noting that we did not identify an optimal temperature that consistently outperformed others across all scenarios in our experiments. Furthermore, we run the feedback iteration loop 4 times. Therefore, in total, we query 5 times in each of the temperatures to generate a patch.
Specifically, we add line number at the beginning of each line of the extracted function block starting from 1.
We add the line numbers with the code block to instruct \chatgpt{} regarding the vulnerable line numbers at the beginning of the prompt.
It is inspired by the best practices for prompt engineering, suggesting adding instructions at the beginning of the prompt~\cite{bestpractice}.

\section{Experimental Setting}
In this section, we first introduce the state-of-the-art approach as our baseline.
And then we provide details of our three research questions.
Last, we describe the evaluation metrics and the dataset used in our experiment.

\subsection{Comparison baseline}\label{sec:baselines}


\tname\ is most related to \codexvr{}, a vulnerability repair technique using zero-shot learning~\cite{pearce2022examining}. Unlike \tname{}, \codexvr{} does not leverage the output of test runs or employ a chain of thought prompt design to improve LLM performance. 
It is worth noting that the original version of \codexvr\ uses the OpenAI's \codex{} model (codex-cushman-001) to produce patches. However, \codex{} has been deprecated\footnote{https://platform.openai.com/docs/guides/code} in March 2023 and, for that reason, we cannot replicate the result of the paper as is~\cite{pearce2022examining}. For a fair comparison, we configure \codexvr\ with \gptthree{}, which is a more recent model developed by OpenAI.
Moreover, there are six different prompts investigated in~\cite{pearce2022examining} as shown in Table \ref{tab:prompts_codexvr}.
Therefore, two configurations potentially impact the approach's performance, i.e., the base model and prompt selection.




\subsection{Research Questions}
\label{sec:research-questions}





Our experiment is designed to answer the following research questions (RQs):

\vspace{1ex}

\noindent \textbf{\rqone{}}


\vspace{1ex}

\noindent\textbf{Rationale}
This question aims to investigate the impact of different prompts and LLMs on the performance of \codexvr{}. In particular, we want to assess if the results do not degrade when using the newer model but for general purpose in a particular task, in our case, i.e., vulnerability repair. It is also important to observe if the opposite occurs ---i.e., if the results of \codexvr\ configured with the new model improve significantly-- as the motivation for proposing a new technique would be weaker in that case.

\noindent\textbf{Setting.} To answer this question we reproduced the setup documented in the \codexvr{} paper ~\cite{pearce2022examining}. 
The only difference between the original experiment and our experiment is the use of \chatgpt{} (instead of \codex{}) as the base LLM. The \codexvr\ paper reported results on various LLMs from the \codex\ family, with \codexModel{} performing best overall. For that reason, and in the interest of space, we restricted the comparison of \chatgpt{} to \codexModel{}.
We evaluate the performance of \codexvr\ on every combination of base model, and all the six prompts proposed in~\cite{pearce2022examining}.
For each combination, we query the model 50 times, as in the \codexvr{} evaluation, and collect the performance on the evaluation metrics~(Section~\ref{sec:metrics}). 

\begin{table}[]
\caption{Prompt templates used in \codexvr~\cite{pearce2022examining}.}
\label{tab:prompts_codexvr}
\vspace{-3ex}
\resizebox{\columnwidth}{!}{%
\begin{tabular}{p{0.8cm}p{9cm}}
\toprule
ID & Description \\ \hline
n.h.      & No Help - deletes the vulnerable code/function body and provides no additional context for regeneration.                                                                                                                  \\ \hline
s.1       & Simple 1 - deletes the vulnerable code/function body and adds a comment ‘bugfix: fixed {[}error name{]}’.                                                                                                                 \\ \hline
s.2       & Simple 2 - deletes the vulnerable code/function body and adds a comment ‘fixed {[}error name{]} bug’.                                                                                                                     \\ \hline
c.        & Commented Code - After a comment ‘BUG: {[}error name{]}’, it includes a ‘commented-out’ version of the vulnerable code/function body followed by the comment ‘FIXED:’. After this it appends the first token of the original vulnerable function. \\ \hline
c.a.      & Commented Code (alternative) - same as c. , but commented in the alternative style for C /* and */ rather than //                                                                                                                        \\ \hline
c.n.      & Commented Code (alternative), no token - same as c.a., but with no ‘first token’ from vulnerable code.                                                                                                                    \\ \bottomrule
\end{tabular}
}\vspace{-5mm}
\end{table}

\vspace{1ex}
\noindent \textbf{\rqtwo{}} 

\vspace{1ex}

\noindent\textbf{Rationale}~
This question aims to assess whether or not our proposed approach outperforms the optimal setting of our comparison baseline, \codexvr, identified through RQ1.

\noindent\textbf{Setting} 
To establish a strong comparison baseline, we use an \emph{idealized optimal version} of \codexvr\ that selects the best prompt and base model identified in RQ1.
We compare such an idealized version of \codexvr\ with \tname{} using the standard metrics from the literature.



\vspace{1ex}
\noindent \textbf{\rqthree{}} 

\vspace{1ex}

\noindent\textbf{Rationale}~ 
The goal of this question is to measure the contribution of different components of \tname\ 
on its performance.

\noindent\textbf{Setting} \tname{} incorporates two features: chain of thought ($C$) and (patch validation) feedback ($F$). We compare the performance of all four feature combinations of \tname: $CF$, $C\neg{}F$, $\neg{}CF$, $\neg{}C\neg{}F$. For $C\neg{}F$, \tname{} does not execute the feedback iteration, i.e., none of the feedback messages is fed from the compiler and test suite. For $\neg{}CF$, \tname{} does not run the CoT block but it performs the feedback iteration. In this setting, none of the trigger sentences have been added at the end of the initial and feedback prompt.

\subsection{Metrics}
\label{sec:metrics}

We use 3 standard metrics from the literature: (1)~percentage of compilable patches and (2)~percentage of plausible patches (i.e., patches that are compilable and also pass the functional and security tests) and (3)~percentage of patches which are semantically equivalent to the ground truth (modulo manual inspection). Functional tests are those available in the test suite. Security tests correspond to the same tests augmented with runtime monitors ---referred to as sanitizers--- to check violations of a security-related property, such as buffer overflows. These monitors (or sanitizers) are automatically introduced by compilers with specific flags (e.g., \CodeIn{-fsanitize=address}~\cite{asan}).



\subsection{Datasets}
\label{sec:dataset}

We evaluate \tname{} on multiple datasets in two popular programming languages, C and Java.

\noindent\textbf{Dataset in C}. We consider ExtractFix~\cite{efbench}, it has been used in previous studies~\cite{pearce2022examining, gao2021beyond}. This dataset consists of \numExamples{} real-world Common Vulnerabilities and Exposures (CVEs) from public open-source projects in C. All the examples include developer patches localized within a single file and have comprehensive test suites.


\noindent\textbf{Dataset in Java}. We also consider two datasets of Java programs (\vjbench~\cite{wu2023effective} and \vulfourj~\cite{bui2022vul4j}) including a total of 50 vulnerabilities with CVE labels. Due to the page limit, please refer to the original papers for more details about these datasets.

\section{Results}
This section reports the results of the experiments to answer the research questions listed in Section~\ref{sec:research-questions}.

\subsection{\rqone{}}
\label{res:rqone}




Table \ref{tab:results-rqone} summarizes the results of \codexvr\ when configured with different LLMs (\codex{} and \chatgpt{}) and prompts\footnote{In interest of space, we provide detailed results on each case in our replication package.}. We use the C dataset in this experiment by following~\cite{pearce2022examining}. We compare the performance of \codexvr\ with two base LLMs and all the six prompts that Pearce et al. proposed in~\cite{pearce2022examining}.


Results show that among all the six prompts, \chatgpt{} performs the best when using the ``c.a.'' prompt, generating 29.6\% plausible patches on average. Differently, \codexvr{} performs the best when using the ``c.'' prompt, yielding an overall average of 9.4\% plausible patches. Overall, \chatgpt{} generates \textbf{+20.2\%} more plausible answer than \codexvr{} in their respective best settings. Furthermore, we find that \chatgpt{} consistently produces more plausible patches in all the prompts. 
These results suggest that \codexvr\ performs better when configured with \chatgpt{} instead of \codex{}.

\begin{table}[t!]
\setlength{\tabcolsep}{2pt}
\caption{\emph{Results of RQ1.} Evaluation of \codexvr{} with different LLM models and prompts. We query each triple of example, model, and prompt \numQueries{} times. The notation $x/y$ indicates $x$ \% plausible patches out of $y$ \% patches that compile. (The higher the values of $x$ and $y$ the better.)
The last column shows whether or not one or more prompts used with a given model report a plausible patch.
}
\label{tab:results-rqone}
\centering
\vspace{-2ex}
\resizebox{1\columnwidth}{!}{%
\begin{tabular}{ccccccccc}
\hline
\multirow{2}{*}{EF\#}     & \multirow{2}{*}{Model}               & \multicolumn{6}{c}{Prompt}                                                                                                     & \multirow{2}{*}{Pass?} \\ \cline{3-8}
                          &                                      & n.h                 & s.1              & s.2                 & c.                    & \textbf{c.a}                  & c.n               &                        \\ \hline
\multirow{2}{*}{Average}  & Codex & 1.8/15.4 & 0.6/10.6 & 1.0/14.4 & 9.4/49.0 & 2.8/55.4 & 5.0/51.8 & 7/10\\
                          & \chatgpt{} &  4.4/24.6 & 3.2/23.2 & 2.4/17.4 & 25.4/44.2 & \textbf{29.6}/\textbf{50.6} & 22.0/62.2 & \textbf{10}/\textbf{10} \\ \hline
\end{tabular}
}\vspace{-3mm}
\end{table}

However, our results also show that the performance of \chatgpt{} is still quite low with only 29.6\% plausible answers in the best case, which is far from desirable.
Based on the column Pass?, We also observe that \codexvr\ with \chatgpt{} can generate plausible for all the cases while \codexvr\ with \codex{} fails to generate plausible patches for 3 cases.
It is worth noting that the count of crosses, by definition, is a lenient performance indicator for techniques.
Overall, our results indicate that simply using a newer LLM is insufficient to repair vulnerabilities effectively.

\noindent
\textbf{Answers to RQ1}: The use of a newer yet general-purpose LLM, i.e., \chatgpt{} instead of \codex{}, does improve the results of \codexvr. However, the performance of \codexvr\ with \chatgpt{} is still far from desirable. This indicates that the adoption of a larger general-purpose language model alone is unlikely to be sufficient to effectively repair vulnerabilities.

\subsection{\rqtwo{}} 
\label{res:rqtwo}

\subsubsection{Evaluation on C dataset}\hfill

\noindent\textbf{Setting}.
Despite the improvement of \codexvr\ when configured with \chatgpt{} instead of \codex{}, results for RQ1 showed that \codexvr\ still performs poorly overall. Motivated by those results, RQ2 compares our proposed technique, \tname\ (Section \ref{technique}), with an \emph{idealized optimal version} of \codexvr\ that is configured with \chatgpt{} and selects the best prompt for a given example, i.e., the prompt that optimizes the metrics for that example (Section~\ref{sec:metrics}).

\begin{table}[]
\caption{Percentage of Compilable and Plausible Patches on C Dataset}
\vspace{-2ex}
\label{tab:results-rqtwo}
\centering
\resizebox{\columnwidth}{!}{%
\begin{tabular}{ccccccc}
\toprule
                     & \multicolumn{3}{c}{\codexvr* on \chatgpt}                                                & \multicolumn{3}{c}{\tname{}}                               \\ \cline{2-7} 
 \multirow{-2}{*}{EF\#} & Compilable\%        & Plausible\%                  & Pass?                & Compilable\%        & Plausible\%                  & \multicolumn{1}{c}{Pass?} \\ \hline
Average              &  55.6 & 33.4 & 10/10 & 90.4 & 63.8 & 10/10 \\\bottomrule
\end{tabular}
}\vspace{-4mm}
\end{table}

\begin{table}[t!]
\caption{Percentage of Correct Patches on C Dataset}
\vspace{-2ex}
\label{tab:results-rqtwo-correctness}
\centering
\resizebox{0.75\columnwidth}{!}{%
\begin{tabular}{ccc}
\toprule
                    & \codexvr* on \chatgpt                                                & \tname{}                              \\ \cline{2-3} 
\multirow{-2}{*}{EF\#}  & \% Correct/ Inspected               & \% Correct/ Inspected       \\ \hline
EF01       & 90 & \textbf{100}   \\
EF02\_01       &  70  & 70         \\
EF08        &  0  &  0        \\
EF15          &  0  &   0        \\
EF17        &  \textbf{100}  &  60     \\
EF18          &  90  & \textbf{100}        \\
EF22          &   0  &   \textbf{70}       \\ \hline
Average          &  50  &  57         \\ \bottomrule
\end{tabular}
}
\vspace{-3ex}
\end{table}

\noindent\textbf{Analysis}. Table~\ref{tab:results-rqtwo} shows the results of comparing \tname\ with the baseline on the C dataset. We use a star (*) to emphasize the use of such an idealized version of \codexvr{}. 
We observe that, both \tname{} and the baseline generates plausible answers in all the cases. \textbf{However, \tname\ can generate 63\% more compilable patches and 91\% plausible patches than the baseline}. Ideally, all generated patches should be compilable. However, results shows that nearly 10\% of generated patches are not.
Furthermore, we manually check the correctness of the generated plausible patches. However, manual validation is known be expensive. Hence, we select 7 (out of 10) cases from C dataset in which \tname{} have produced more than 10 plausible patches.
For each case, we randomly select at most 10 patches among the generated plausible patches.
One labeler carefully examines the selected patches and then discusses the result with the second labeler to derive the final decision on the correctness.
Table~\ref{tab:results-rqtwo-correctness} presents the result of manual inspection. We find that \textbf{\tname{} produces 14\% more correct patches compared to the baseline}. However, our results indicate that there is still significant room for improving patch correctness.


\subsubsection{Evaluation on Java dataset}\hfill

\noindent\textbf{Settings}. We evaluate \tname{} on the \vjbench~\cite{wu2023effective} and \vulfourj~\cite{bui2022vul4j} datasets of Java vulnerabilities. We follow a similar experimental procedure that
Wu et al.~\cite{wu2023effective} use to evaluate the ability of the LLMs from the Codex family to generate correct patches. Wu et al. report that the davinci-002 model (we refer to it \vjfix) performed the best, therefore we reuse the same configuration setting. Specifically, we run \tname{} by querying the LLM with a temperature 0.6 and generating 10 patches for each vulnerability. It is worth noting, however, that we run our pipeline once --for the sake of time and computational cost-- whereas authors of \vjfix\ run their pipeline 25 times, taking averages of the metrics of interest.

\begin{table}[t!]
\centering
\caption{Performance of \tname{} on fixing Java vulnerabilities. The values under `Plausible' column denotes the number of plausibly fix vulnerability (at least one patch that passes the test cases) and the values under `Correct' denotes the number of correctly fix vulnerability (at least one patch that is semantically equivalent to human patch)}
\label{tab:results-vjbench}
\vspace{-2ex}
\resizebox{0.8\columnwidth}{!}{%
\begin{tabular}{ccccc}
\toprule
& \multicolumn{2}{c}{\vjfix} & \multicolumn{2}{c}{\tname}  \\ 
& Plausible & Correct & Plausible & Correct \\
\midrule
VjBench (15)  & 4.6  & 4.0 & \textbf{6.0}  & \textbf{5.0} \\
Vul4j (35)  & 10.9  &  6.2 & \textbf{14.0} & \textbf{9.0} \\ 
\midrule
Total (50) & 15.5 & 10.2 & \textbf{20.0} & \textbf{14.0} \\ 
\hline
Compilation Rate (\%)  & \multicolumn{2}{c}{79.7}  & \multicolumn{2}{c}{\textbf{86.0}}  \\ 
\bottomrule
\end{tabular}
}
\vspace{-4mm}
\end{table}

\noindent\textbf{Analysis}.
Table~\ref{tab:results-vjbench} shows the results comparing \vjfix\ and \tname. We observe an increase from 79.7\% to 86\% (6.3\% gain) in terms of compilation rate and from 15.5 to 20.0 (9.0\% gain) in terms of plausible patch when using \tname\ instead of the baseline.
Finally, considering the number of cases in which a semantically correct patch is generated (out of 50), we observe an increase from 10.2 to 14 (7.6\% gain). As usual, semantic correctness is based on manual inspection where one of the authors evaluates and discusses with other authors if the patch is behavioural-equivalent to the human patch. It is worth noting that there are 4 cases where \tname\ generates plausible patches for which we were unable to decide correctness because of the complexity of the code. 
For instance, the human patch for fixing vulnerability Vul4J-8 adds two new conditions to break one loop.\footnote{Human patch: \url{https://github.com/apache/commons-compress/commit/4ad5d80a6272e007f64a6ac66829ca189a8093b9}} \tname{} generates a plausible patch for this vulnerability by adding different loop condition.\footnote{Diff between original and patch: \url{https://www.diffchecker.com/7aePjrvI/}} Although the patch seems logically correct, it is hard to fully ascertain without domain knowledge. We further discuss this on Section~\ref{sec:lessons}

\noindent
\textbf{Answers to RQ2}: 
\tname\ consistently outperforms the baselines on both C and Java datasets in terms of compilable, plausible and correct patch rate. However, generating the correct patch remains challenging.


\subsection{\rqthree{}}
\label{res:rqthree}


Table~\ref{tab:results-rqthree} shows the results of the four different combinations of these two components.
This RQ considers the percentage of plausible patches as our main evaluation metric.
We observe that the baseline without both Chain-of-Thought and Feedback produces 23\% plausible patches. 
Column ``Plausible'' under ``Without Chain-of-Thought'' and ``Without Feedback'' shows that information.
We construct the prompt for this combination by removing the chain-of-thought and the feedback components from \tname{}. 


\begin{table}[]
\centering
\setlength{\tabcolsep}{2pt}

\caption{\emph{Results of RQ3.} Impact of \tname's components. The leftmost combinations shows the impact of \tname's prompt without any of the components (CoT and Feedback). The rightmost combination shows the aggregated impact of all component. The two combinations in the middle show the impact of each component in isolation. }
\vspace{-3ex}
\label{tab:results-rqthree}
\resizebox{1\columnwidth}{!}{%
\begin{tabular}{ccccccccc}
\toprule
                       & \multicolumn{4}{c}{w/o Chain of Thought}                                                                               & \multicolumn{4}{c}{w/ Chain of Thought}                                                                                  \\ \cline{2-9} 
                       & \multicolumn{2}{c}{w/o Feedback}                       & \multicolumn{2}{c}{w/ Feedback}                             & \multicolumn{2}{c}{w/o Feedback}                       & \multicolumn{2}{c}{w/ Feedback}                             \\ \cline{2-9} 
\multirow{-3}{*}{Id} & Compilable           & Plausible                           & Compilable           & Plausible                              & Compilable           & Plausible                           & Compilable           & Plausible                              \\ \hline
EF01                   & 60                   & \cellcolor[HTML]{F3F3F3}0           & 96                   & \cellcolor[HTML]{DEE7D9}16             & 76                   & \cellcolor[HTML]{96C084}68          & 88                   & \cellcolor[HTML]{86B770}80             \\
EF02\_01               & 100                  & \cellcolor[HTML]{E9EDE6}8           & 100                  & \cellcolor[HTML]{DEE7D9}16             & 96                   & \cellcolor[HTML]{A7C998}56          & 100                  & \cellcolor[HTML]{8EBC7A}74             \\
EF07                   & 80                   & \cellcolor[HTML]{E9EDE6}8           & 92                   & \cellcolor[HTML]{B7D2AB}44             & 52                   & \cellcolor[HTML]{D8E4D3}20          & 88                   & \cellcolor[HTML]{A7C998}56             \\
EF08                   & 56                   & \cellcolor[HTML]{EEF0ED}4           & 88                   & \cellcolor[HTML]{C8DBBF}32             & 60                   & \cellcolor[HTML]{D8E4D3}20          & 76                   & \cellcolor[HTML]{9CC38B}64             \\
EF09                   & 92                   & \cellcolor[HTML]{CDDEC6}28          & 100                  & \cellcolor[HTML]{C8DBBF}32             & 68                   & \cellcolor[HTML]{C2D8B8}36          & 96                   & \cellcolor[HTML]{7BB163}88             \\
EF15                   & 92                   & \cellcolor[HTML]{F3F3F3}0           & 100                  & \cellcolor[HTML]{F3F3F3}0              & 80                   & \cellcolor[HTML]{EEF0ED}4           & 100                  & \cellcolor[HTML]{BDD5B2}40             \\
EF17                   & 92                   & \cellcolor[HTML]{75AE5D}92          & 96                   & \cellcolor[HTML]{70AB56}96             & 88                   & \cellcolor[HTML]{C8DBBF}32          & 96                   & \cellcolor[HTML]{75AE5D}92             \\
EF18                   & 96                   & \cellcolor[HTML]{70AB56}96          & 100                  & \cellcolor[HTML]{6AA84F}100            & 92                   & \cellcolor[HTML]{86B770}80          & 100                  & \cellcolor[HTML]{6AA84F}100            \\
EF20                   & 100                  & \cellcolor[HTML]{F3F3F3}0           & 100                  & \cellcolor[HTML]{EEF0ED}4              & 76                   & \cellcolor[HTML]{F3F3F3}0           & 92                   & \cellcolor[HTML]{EEF0ED}4              \\
EF22                   & 96                   & \cellcolor[HTML]{F3F3F3}0           & 100                  & \cellcolor[HTML]{E3EAE0}12             & 56                   & \cellcolor[HTML]{CDDEC6}28          & 68                   & \cellcolor[HTML]{BDD5B2}40             \\ \hline
Average    &           86 & 23 & 97 & 35 & 74 & 34 & 90 &  64 \\ \hline
\end{tabular}
}
\vspace{-5ex}
\end{table}

Adding chain-of-thought to the baseline increases the rate of plausible patches generated from 23\% to 34\% (+47\% gain). This result shows the positive impact of the chain-of-thought prompt.
Adding feedback to the baseline increases the rate of plausible patches from 23\% to 35\% (+52\% gain). When combining chain of thought with feedback, the rate of plausible patches increases to 64\%, reflecting the importance of these two components to \tname.

Results show that combining Chain-of-Thought with Feedback (i.e., our method \tname{}) yields the best result. In this setting, the percentage of plausible patches increases to 64\%. Moreover, the distribution of green cells across the table suggests that Feedback and Chain-of-Thought synergistically cooperate. In four of the cases (EF02\_01~\cite{EF0201}, EF08~\cite{EF08}, EF09~\cite{EF09}, and EF15~\cite{EF15}), the rate of plausible answers in the full-fledged combination is higher than the sum of the rates from the combinations with a single component.


\noindent
\textbf{Answers to RQ3}: 
Each component of \tname is essential to its performance and the full-fledged combination achieves the best or equally the best performance, suggesting that the components synergistically cooperate.

\section{Discussion}

\subsection{Lessons and Implications}
\label{sec:lessons}

\begin{figure}[]
    \centering
\includegraphics[width=\columnwidth]{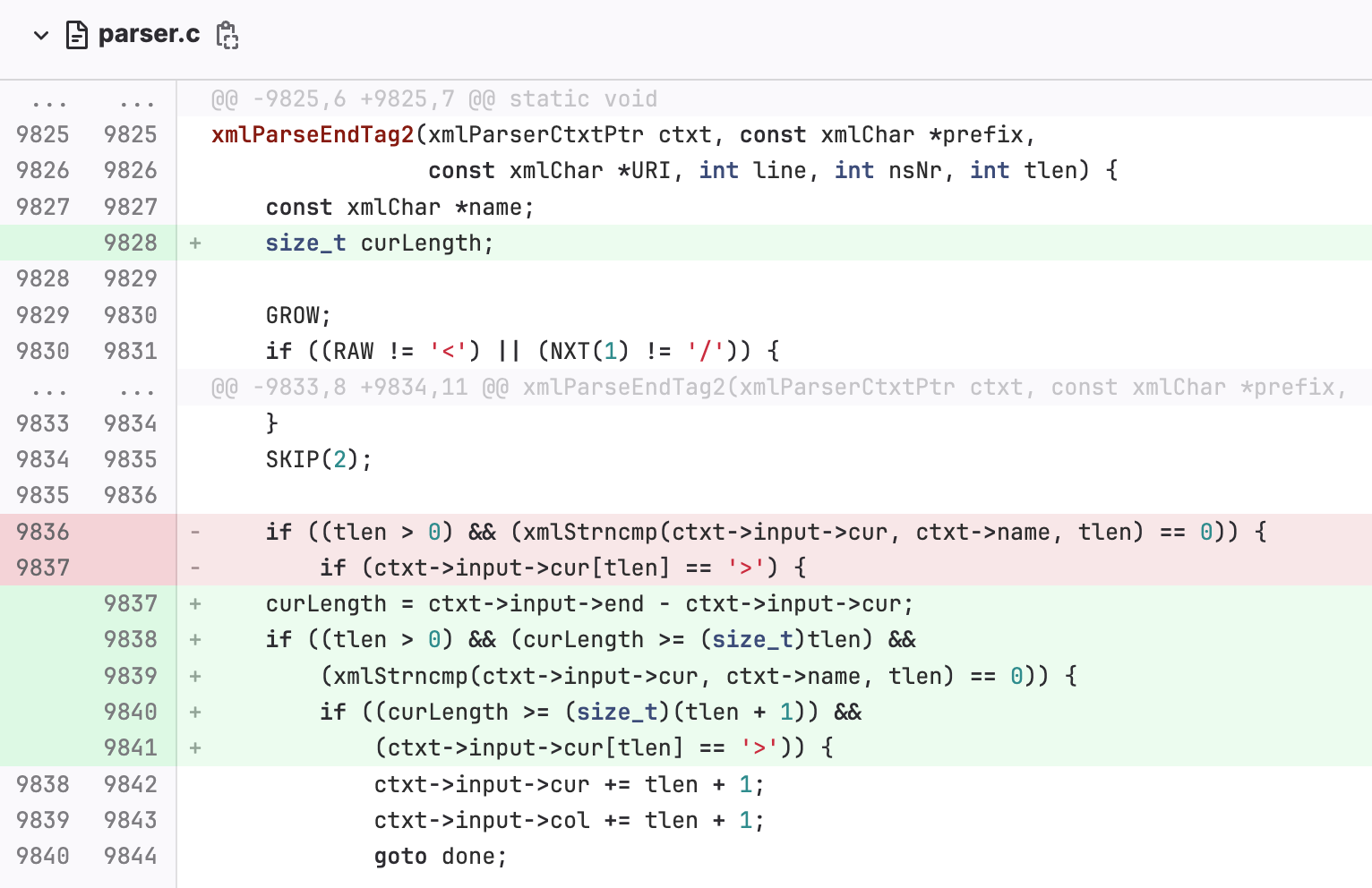}
    \vspace{-4mm}
    \caption{A Complex Patch for EF15 \textit{out-of-bounds} vulnerability}
    \label{fig:ef15}
    \vspace{-5mm}
\end{figure}

\noindent\textbf{Lesson \#1 The use of a reason-feedback mechanism can enhance the ability of Large Language Models (LLMs) in repairing vulnerabilities, though addressing complex vulnerabilities remains a significant challenge}. This observation is supported by the result of RQ3 (Section~\ref{res:rqthree}). We found that combining a chain-of-thought approach for reasoning  with error messages from tests for feedback can improve the performance of LLMs. Specifically, our manual verification confirmed all plausible patches randomly selected for vulnerabilities EF01~\cite{EF01} and EF18~\cite{EF18} are correct. These vulnerabilities, categorized as \textit{out-of-bounds} and \textit{null pointer dereference}, respectively, are often resolved by adding a simple conditional check. \textbf{However, the complexity of a vulnerable code fix cannot be determined only by the corresponding vulnerability category}. For example, in our experiments, EF15~\cite{EF15} is also categorized as \textit{out-of-bounds} vulnerability. However, as shown in Table~\ref{tab:results-rqtwo-correctness}, none of the randomly selected plausible patches were found to be correct upon manual examination. 
Figure~\ref{fig:ef15}
presents the human-generated patch for EF15.
To fix this vulnerability, it requires a significant amount of code change. A new variable, \CodeIn{curLength}, is created and initialized. That change requires a deeper understanding about the data structure \CodeIn{xmlParserCtxtPtr}.

\vspace{-2mm}
\begin{tcolorbox} [boxsep=1pt,left=2pt,right=2pt,top=1pt,bottom=1pt]\textbf{Implication \#1}: 
Context is the key for LLM in vulnerability repair. Additional information, such as vulnerability description and test cases outcomes, is helpful but insufficient for fixing complex vulnerabilities.
\end{tcolorbox}
\vspace{-2mm}

\noindent\textbf{Lesson \#2 Fixing vulnerabilities that are inherently linked to project design poses a significant challenge}. An example of this is EF08 (CVE-2017-7601~\cite{CVE_2017_7601}), a vulnerability arising from LibTIFF's handling of undefined behavior ``shift exponent too large for 64-bit type long.'' Remote attackers can exploit this vulnerability to create a denial of service attack (making the application to crash) or possibly other unspecified behavior via a crafted image.
A feasible solution to mitigate this vulnerability involves implementing a condition that checks whether the shift exponent size surpasses a certain bit threshold.
Aligning with standard JPEG specifications, thresholds of 8 or 12 bits per sample are often considered. However, an important part of the human patch for EF08~\cite{EF08} is adding a line of code \CodeIn{if(td->td\_bitspersample > \textbf{16})},
the threshold is intentionally set to 16 bits. The choice of this threshold value reflects a design decision by the developers, influenced by factors such as their application's specific requirements, the anticipated range of TIFF files to be processed, and their strategy for balancing flexibility with safety and adherence to standards. This vulnerability underscores the complex challenge of addressing security issues that necessitate consideration of the project's design context. Moreover, we find that the key reason of \tname{} fails to generate a correct patch is because it is ``misled'' by the vulnerability description, i.e., ``shift exponent too large for 64-bit type long.'' Patches generated by \tname{} set the threshold to 64 instead 16 bits. Our finding reveals the challenge that \textbf{a deeper understanding on project/environment-related information beyond vulnerability description may be required for repairing complex vulnerabilities}.

\vspace{-2mm}
\begin{tcolorbox} [boxsep=1pt,left=2pt,right=2pt,top=1pt,bottom=1pt] \textbf{Implication \#2}: 
Domain knowledge (e.g., project design) is critical to repair vulnerabilities. However, extracting this information is challenging. Investigating these aspects through LLMs presents a promising avenue for future research.
\end{tcolorbox}
\vspace{-2mm}




\noindent\textbf{Lesson \#3
The pressing need for a comprehensive and high-quality dataset for vulnerability repair is evident}. Constructing such a dataset is a complex task. It should encompass both functionality and security tests to ensure thoroughness. Additionally, the dataset must be designed for easy replication of the execution environment. For instance, in our work with the ExtractFix dataset, Docker was utilized to establish a consistent executable environment. Furthermore, comprehensive documentation of the dataset's usage is crucial. Currently, locating datasets in vulnerability repair with \textit{functionality-and-security test offered}, \textit{well-documented}, \textit{reproducible}, \textit{well-maintained}, and \textit{scalable} is a formidable challenge. These difficulties are similarly highlighted in a recent study from Croft et al.~\cite{croft2023data}. Therefore, addressing these challenges is essential for advancement in this field and significantly reducing the engineering burden for researchers.


\vspace{-2mm}
\begin{tcolorbox}[boxsep=1pt,left=2pt,right=2pt,top=1pt,bottom=1pt] \textbf{Implication \#3}: 
High-quality datasets of vulnerability repair are rare. We urgently advocate more community involvement in developing reliable, reproducible and easy-to-use datasets with minimum human efforts.
\end{tcolorbox}
\vspace{-2mm}


\noindent\textbf{Lesson \#4 Fully manual patch validation for vulnerable code sometimes can be extremely difficult}. This process, while crucial for distinguishing plausible from correct patches, demands highly skilled labelers with not only extensive programming experience but also a profound understanding of specific vulnerabilities. Such expertise makes it difficult to find suitable candidates. Moreover, comprehending vulnerabilities, often subtle and deliberately crafted by attackers, tends to be more complex than understanding general bugs. Additionally, while human-generated patches can offer insights into the logic of fixes, they sometimes present their own difficulties in interpretation. For instance, 
the human patch for the \textit{Divide By Zero} vulnerability (EF09)~\cite{EF09}
exemplifies this. It needs deep comprehension of the code context, including specific variables (\CodeIn{horizSubSampling} and \CodeIn{vertSubSampling}) and function calls (\CodeIn{usage(-1)}). This complexity adds another layer of challenge to the manual validation process.

\vspace{-2mm}
\begin{tcolorbox}[boxsep=1pt,left=2pt,right=2pt,top=1pt,bottom=1pt]\textbf{Implication \#4}: 
Manually inspecting the correctness of a generated security patch is expensive. 
Utilizing large language models for semi-automatic patch inspection offers potential promise for achieving an optimal balance between accuracy and scalability.
\end{tcolorbox}
\vspace{-2mm}

\subsection{Threats to Validity}
\label{sec:threats}

The first threat to validity comes from the correctness of the plausible patches compared with the reference developer patch. To address this threat, we carefully examined and discussed each patch following prior work~\cite{xia2023keep,pearce2022examining,jiang2021cure,xia2023automated}. However, manually measuring the correctness of the plausible patches is expensive. Some of the plausible patches have big changes and understanding the correctness of these patches requires domain specific knowledge. We sought second opinions and conservatively labeled patches as non-equivalent, when needed. 
The second threat to validity comes from the potential data leakage issue of developer patches being part of the original training data of ChatGPT. ChatGPT is not open-sourced and can only be accessed through API, so we cannot access the exact training data used. 
This improvement gain is doubtful to be by memorizing training data.
However, our approach and the baseline (RQ2) are based on the same version of \chatgpt{} while our approach outperforms the baseline.

\vspace{-2mm}
\section{Related Work}
Researchers have proposed vulnerability repair techniques by leveraging recent advances in deep learning. For instance, Chen et al.~\cite{chen2022neural} proposed VRepair based on transformer and transfer learning. They showed that pre-training improves repair performance compared to training from scratch. Fu et al.~\cite{fu2022vulrepair} propose VulRepair, a technique that uses sub-word tokenization and pre-training for vulnerability repair, and show that that VulRepair outperforms VRepair.
Huang et al.~\cite{huang2022repairing} applied pre-trained models for vulnerability repair to overcome the shortcomings of learning-based APR techniques. They demonstrated that these pre-trained models outperform learning-based
APR techniques (e.g., CoCoNut~\cite{Lutellier2020CoCoNuTCC} and DLFix~\cite{Li2020DLFixCC}) and more data-dependent features help repair complex vulnerabilities.
More recently, Wu et al.~\cite{wu2023effective} proposed a Java vulnerability dataset, VJBench, by enhancing Vul4J~\cite{bui2022vul4j}. They demonstrate that existing LLMs and APR models fix very few Java vulnerabilities. These works use large language models trained for code-related tasks.
In contrast to prior work, this paper focuses on assessing effectiveness --for the task of vulnerability repair-- of using a general-purpose LLM (e.g., \chatgpt{}) in combination with techniques that have been shown effective to reduce the LLM-to-code semantic gap.

\vspace{-2mm}
\section{Conclusion} 
This paper examines the idea of combining reasoning and patch validation feedback for vulnerability repair. The idea was originally proposed by Xia et al~\cite{xia2023keep} in the general context of automated program repair. We present \tname{}, a method that queries the LLM first to reason about the code's specific vulnerability and then to respond with a repair patch using a chain of thought prompt. Furthermore, \tname{} iteratively refines the incorrect patch using error messages produced by the compiler and test suite as feedback to the LLM.
We evaluate \tname{} on multiple datasets for Java and C.
The results we obtained indicate that the idea of using reasoning and patch validation is still valid in the context of vulnerability repair, however, we find that the performance of LLM for this task still is far from desirable.
This paper discusses the lessons we learned from this study and highlights the challenges for advancing LLM-based vulnerability repair.







\vspace{1ex}\noindent\textbf{ACKNOWLEDGMENTS.}~
This work was supported and funded by the National Science Foundation Grant No. 2026928. Any opinions expressed in this material are those of the authors and do not necessarily reflect the views of any of the funding organizations.


\balance
\bibliographystyle{ACM-Reference-Format}
\bibliography{ref,refsDL}










\end{document}